# Hybrid plasmonic nanostructures based on controlled integration of MoS$_2$ flakes on metallic nanoholes

*Denis Garoli[1*], Dario Mosconi[2], Ermanno Miele[1], Nicolò Maccaferri[1], Matteo Ardini[1], Giorgia Giovannini[1], Michele Dipalo[1], Stefano Agnoli[2] and Francesco De Angelis[1*]*

[1] Istituto Italiano di Tecnologia, via Morego 30, I-16163, Genova, Italy.

[2] Dipartimento di Chimica, Università degli Studi di Padova, Via Marzolo 1, 35131, Padova, Italy.

\* Corresponding author: Dr. Denis Garoli, denis.garoli@iit.it; Dr. Francesco De Angelis, francesco.deangelis@iit.it

## Abstract

Here, we propose an easy and robust strategy for the versatile preparation of hybrid plasmonic nanopores by means of controlled deposition of single flakes of MoS$_2$ directly on top of metallic holes. The device is realized on silicon nitride commercial membranes and can be further refined by TEM or FIB milling to achieve the passing of molecules or nanometric particles through a pore. Importantly, we show that the plasmonic enhancement provided by the nanohole is strongly accumulated in the 2D nanopore, thus representing an ideal system for single-molecule sensing and sequencing in a flow-through configuration. Here, we also demonstrate that the prepared 2D material can be decorated with metallic nanoparticles that can couple their resonance with the nanopore resonance to further enhance the electromagnetic field confinement at the nanoscale level. This method can be applied to any gold nanopore with a high level of reproducibility and parallelization; hence, it can pave the way to the next generation of solid-state nanopores with plasmonic functionalities. Moreover, the controlled/ordered integration of 2D materials on plasmonic nanostructures opens a pathway towards new investigation of the following: enhanced light emission; strong coupling from plasmonic hybrid structures; hot electron generation; and sensors in general based on 2D materials.

Nanopore technology is the core of third-generation sequencing, and solid-state nanopores are now one of the main topics of research in single-molecule sensing. To produce solid-state nanopores, a valid alternative to the currently used commercial biological nanopores, as well as the current nanofabrication methods and materials, must be developed. To date, one of the most advanced generations of solid-state nanopores is represented by 2D materials. The atomically thin nature of graphene and other materials, such as transition metal dichalcogenides (TMDCs) ($MoS_2$, $WS_2$, etc.)[1] make them ideal translocation membranes for high-resolution, high-throughput, single-molecule sequencing based on nanopores[2–12]. Electrical measurements are the main approach for single-molecule sequencing by means of nanopores, but readout schemes that rely on optical spectroscopy can be envisioned[13-15]. Within this context, a plasmonic nanopore[15] represents an intriguing tool for enhancing the signal-to-noise from the optical signal via the electromagnetic field enhancement that can be generated by engineered metallic nanostructures. The integration of a 2D material with plasmonic nanostructures leads to a new generation of hybrid nanopores[16]. In fact, the development of a hybrid 2D-plasmonic nanoarchitecture that efficiently combines the benefits from a plasmonic field enhancement with the intrinsic in-plane electric field localization from atomically thin materials would represent a step towards the next generation of hybrid nanopores.

The preparation of a nanopore in an atomically thick layer of 2D material remains a challenging task and requires the deposition of single layers on top of a larger solid-state pore/membrane with a successive step involving a high-resolution electron beam sculpting/drilling process[9] that often suffers from process variability, precluding the platform from being scalable. Moreover, to prepare a hybrid plasmonic / 2D material pore, the integration of metallic nanostructures must be achieved in close proximity to the 2D material pore. The pioneering work of Nam et al.[16] on a hybrid nanopore involved the use of photothermal sculpting to create a nanopore in a graphene membrane with a nearby gold nanoparticle acting as an optical antenna. In their paper, a significant enhancement in fluorescence was detected during single-molecule DNA translocation though the nanopore, highlighting the potential of hybrid systems. Here, we propose an easy and robust strategy for the versatile preparation of hybrid plasmonic nanostructures by means of controlled deposition of single- or few-layer flakes of $MoS_2$ directly on top of metallic holes. This method can be applied to any gold nanohole (with 2D or 3D geometries, as will be demonstrated) and can pave the way to the next-generation fabrication of hybrid systems. Compared to the more frequently used graphene, MoS$_2$ presents several advantages in nanopore applications and has been recently proposed for use in single-molecule sequencing[4,5,11]. The presence of local defects (-S vacancies) on the MoS$_2$ layer is used here to anchor the material locally on metallic nanoholes by means of chemical functionalization via dithiol molecules. Thiol conjugation of MoS$_2$ has been explored in a few recent works[17-19]; to the best of our knowledge, this work represents the first example of a deposition method based on that process. A pictorial representation of the process is illustrated in Fig. 1 (Top Panel). The preparation of MoS$_2$ flakes is based on chemical exfoliation[20-22], as described in the SI; Figures 1(b)-(d) show examples of the ordered deposition of single flakes on top of 2D and 3D gold holes. The method used for the deposition is based on the conjugation between a gold (or another noble metal) surface and a dithiol-terminated organic chain as well as the same conjugation between the MoS$_2$ flake and the dangling –SH group of the same molecule (Fig. 1). In particular, to perform a controlled deposition of MoS$_2$ over metallic holes, we used a 1,12-dodecanedithiol molecule as a linker between the gold surface and local –S vacancies in MoS$_2$ flakes. The protocol of functionalization is the following (illustrated in Fig. 1-Top Panel): 1) a 1 mM solution of dithiol is prepared in EtOH; 2) the sample to be deposited is first cleaned in oxygen plasma for 60 seconds to facilitate the process; 3) the plasmonic holes are prepared on a Si$_3$N$_4$ membrane, and only one side of the substrate is covered with metal (see the SI for details on the nanohole fabrication process); 4) as we expect the thiol deposition to occur only on the metal in contact with the solution, to functionalize only the holes, we put the sample with the metallic side in contact with a MoS$_2$ suspension in EtOH, i.e., we keep the sample floating to avoid the complete wetting of the sample; 5) at the same time, we spot on the dry side, opposite to the one we want to decorate with MoS$_2$, 10 µL of 1,12-dodecanedithiol diluted in EtOH; 6) after a few seconds, the drop of dithiol starts to dry; and 7) the sample is rinsed in EtOH, and the controlled deposition is achieved. This method can be used on every nanostructure involving a metallic nanohole; in our case, we demonstrate that the link between the two materials can be achieved with high reproducibility, both in a flat metallic hole and in 3D hollowed antennas. To control the % of coverage of the holes and the quantity of deposited flakes in terms of number of layers, the most critical parameter is the time of incubation. In our case, 20 seconds of incubation leads to a high percentage of single-layer flakes deposited over large nanohole array. In fact, as will be reported later, over 80 % of metallic holes present in the array can be covered with MoS$_2$. The optimization of the preparation of a MoS$_2$ batch (see the SI), allows one to obtain a single-layer deposition almost over the different pores. This deposition can be demonstrated by Raman measurements where the discrimination among single- or few-layer flakes is possible[23]. As illustrated in Fig. 1(a), (b) and (c), the deposition over the 2D nanopores results in small flakes covering the pores with dimensions spanning from 200 nm up to 500 nm. In the case of the 3D structures (Fig. 1(d), (e) and (f)), the flake deposition results in a partially covering layer that crinkles around the metal in many different ways, ranging from small flakes covering only the top hole to large flakes wrapping the 3D body of the structure. Note that, although it is beyond the scope of this work, the obtained structures can find several interesting applications in all the present fields of research in which 2D materials are the core. For example, the controlled/ordered integration of 2D materials on plasmonic nanostructures can pave the way to new investigations on enhanced light emission from TMDCs[24-31], strong coupling from plasmonic hybrid structures[32], hot electron generation [33-34], and sensors in general based on 2D materials[35-40].

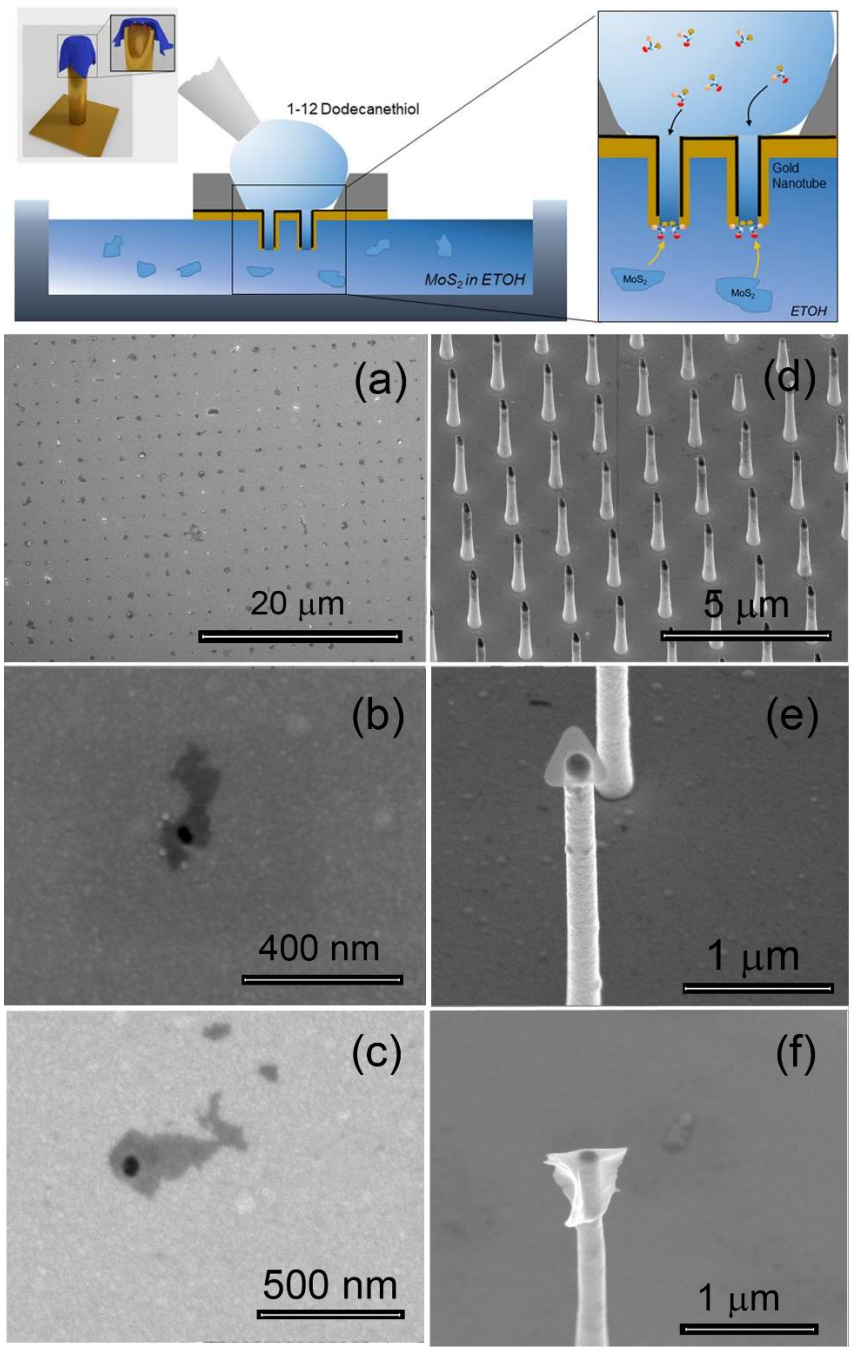

*Figure 1*. SEM micrographs of MoS$_2$ flakes deposited onto an array of plasmonic nanoholes. (top panel) Illustration of the concept for controlled deposition of MoS$_2$ flakes over metallic holes; (a) top view over large flat gold holes array; (b)(c) detail of a single-layer flake on a 2D pore; (d) tilted view over large array of 3D antennas covered with MoS$_2$ flakes; (e)(f) details of the MoS$_2$ flakes deposited onto an antenna.

Next, we investigate the plasmonic properties of our archetypical 3D structure integrated with MoS$_2$ flakes by means of finite element method (FEM) simulations using the RF Module in Comsol Multiphysics and taking into account the geometry that can be fabricated using our method [41,42]. The phenomena that will be illustrated resemble the phenomena we expect to observe in a flat metallic pore integrated with the 2D flake (see SI). The optical properties of MoS$_2$ flakes can be simulated considering the experimental optical constants obtained by Zhang et al. [43] for a single-layer MoS$_2$ film in the spectral range of interest. First, we consider a hollow 3D antenna with optimized dimensions (height, diameter and hole radius) for field confinement at the top area at wavelength of 633 nm. We used this antenna structure for the Raman spectroscopy test reported below (details on fabrication are reported in the SI). At this wavelength, it is possible to enhance the electromagnetic field at the top of our structure by almost two orders of magnitude. As shown in Fig. 2(b), the same significant field enhancement can also be obtained in the case of a hollow antenna with a monolayer of MoS$_2$ covering the hole, as in the experiment. Note that the presence of a top MoS$_2$ layer does not significantly change the field distribution in the region of interest. Because we are investigating the fabrication of a nanopore into a MoS$_2$ monolayer, a 5 nm hole has been simulated as well. As seen from Fig. 2(c), the presence of such a hole in the high-index MoS$_2$ layer on top of the plasmonic antenna induces a strong field confinement and intensity enhancement (by a factor of 50) inside the nanopore. Note that this high field enhancement is not achievable if no plasmonic element is present, i.e., if we consider a hollow antenna without the gold coverage, we have the field confinement and a small enhancement (by a factor of 3) inside the nanopore (see the SI), whereas with the gold, we increase this enhancement by at least one order of magnitude.

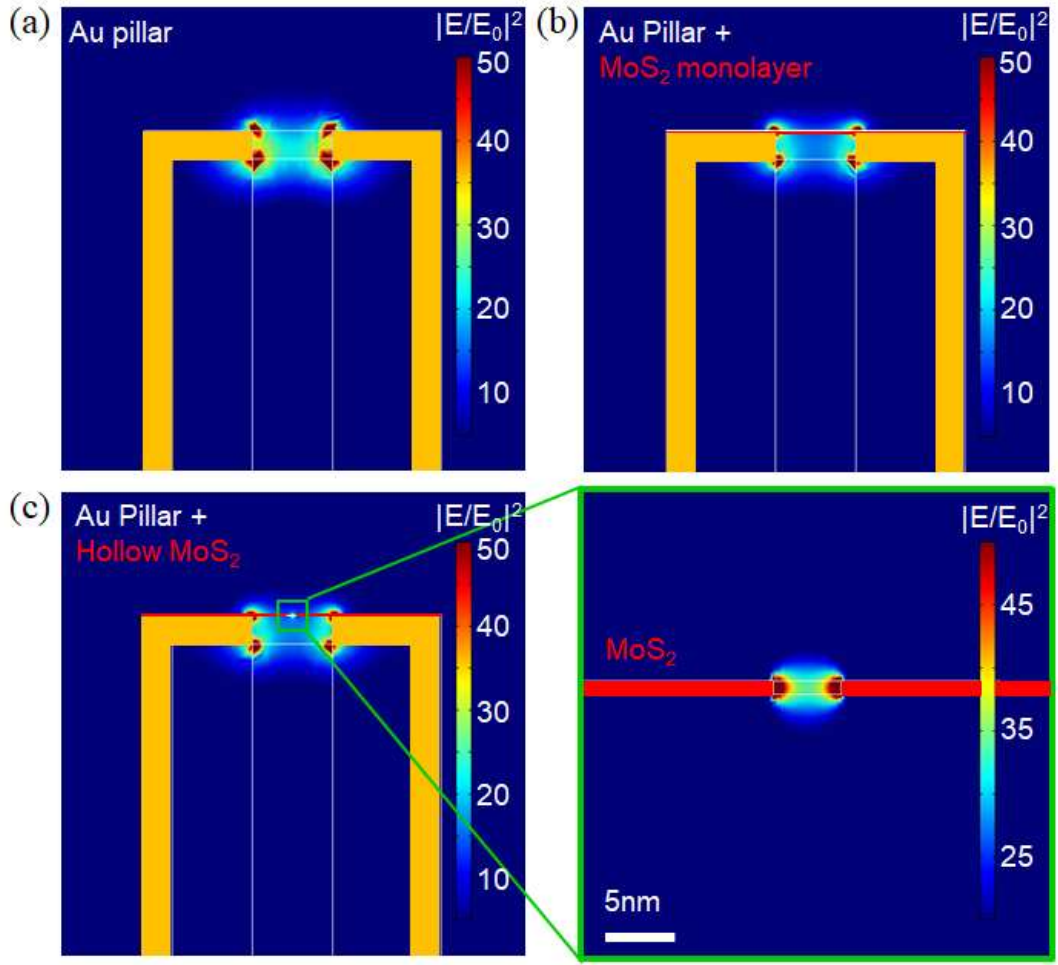

*Figure 2*. (a) Electric field intensity distribution of a gold nanopillar made of a 300 nm wide and 400 nm thick dielectric (S1813 optical resist[41,42]) structure covered with 35 nm of gold at $\lambda$=633 nm. (b) Electromagnetic field distribution with a disk with a $MoS_2$ monolayer on top of the pillar. (c) Left-panel: same as in (b) but with a circular nanohole 5 nm wide in the $MoS_2$ monolayer; right-panel: electric field intensity distribution in the $MoS_2$ nanohole.

As is well known, $MoS_2$ is a layered material; thus, to obtain processable flakes, it must be exfoliated by breaking the van der Waals interactions between the layers, a process that causes high stress to the material in all the preparation methods developed until now[20-22].

Considering that each $MoS_2$ single-layer nanosheet consists of molybdenum atoms sandwiched between two planes of sulfur anions, it is reasonable to suppose that several defects are present on the surface where vacancies in –S bonds are expected[44]. In principle, these unsaturated bonds can represent not only a favorite site for the thiol conjugation used for the deposition but also a site of nucleation for gold nanoparticles, hence allowing the decoration of the flakes. In addition, if we consider the higher sputtering rate of S respect to Mo atoms, if a $MoS_2$ layer is drilled with an ion/electron beam, then we can expect to have extra edges where free –S links (or partially unsaturated S anions) may be present. Moreover, these latter free bonds can be used as nucleation sites for metallic nanoparticle growth or deposition.

Consequently, here we extend our simulations considering the case of a 5-nm gold nanoparticle (AuNP) in close proximity to a 5-nm pore prepared in the $MoS_2$ flake. As reported below, this case resembles well the experimental case where the feasibility of this fabrication is demonstrated.

Figure 3 illustrates the effect as obtained from our FEM simulations. As expected, the presence of a metallic nanoparticle on the edge of the $MoS_2$ nanopore strongly distorts the field confinement, leading to significant additional enhancement due to coupling between the plasmonic nanopore and the resonating AuNP. Moreover, switching the polarization appears to possibly modulate this coupling phenomenon and hence the final enhancement. In fact, as illustrated in Fig. 3(a) and 3(b), once the polarization of the incident light is oriented along the AuNP and the hole, the field reaches a value of up to 90 at the pore exit, whereas in the cross polarized configuration (Fig. 3(c) and 3(d)), this effect cannot be observed. This result suggests a possible means to switch the system based on this effect. For example, this switching could be interesting for applications where single molecules pass through the pore for sensing based on enhanced spectroscopy, such as SERS or metal enhanced fluorescence (MEF)[14,15], both of which are now of great interest for sequencing applications.

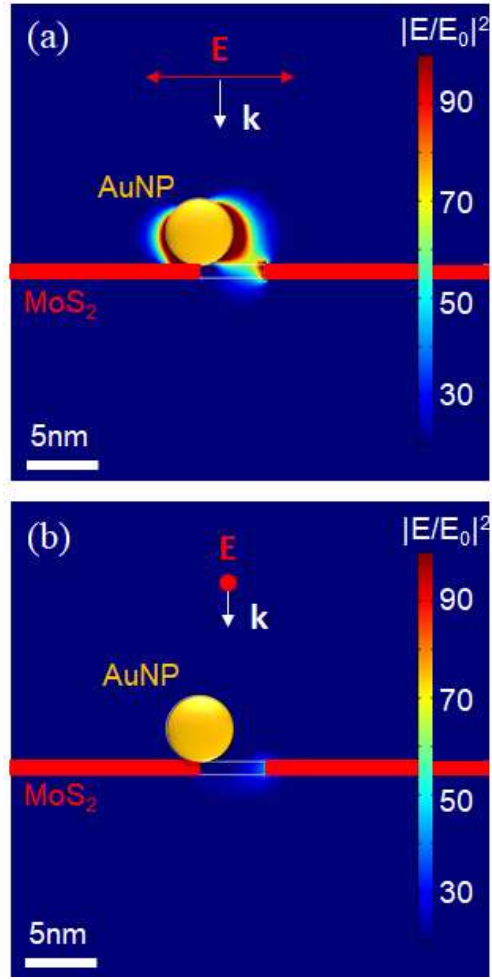

***Figure 3**. (a) Electric field intensity distribution at the nanopore when the incident electric field is parallel to the AuNP-nanopore axis. (b) Electric field intensity distribution at the nanopore when the incident electric field is perpendicular to the AuNP-nanopore axis.*

From the perspective of the fabrication point, as is well known, a single layer of $MoS_2$ is approximately 0.7 nm thick, and a nanopore can be easily prepared by means of focused electron beam exposure (using TEM)[4, 5]. Here, we first verify that the deposited $MoS_2$ flakes can be sculptured by means of a 100 keV focused electron beam (HRTEM Fig. 4(a) and (b)). However, TEM sculpturing is a very time-consuming and expensive procedure and can be performed only on suitable small and fragile substrates. Alternative strategies for narrow nanopore fabrication represent important contributions to the nanopore topic[2, 45, 46]. Here, we report the preparation of a sub-10-nm hole prepared in a $MoS_2$ flake by means of FIB milling at low current (4.4 pA) with a single-pass exposure. The ability of the FIB microscope to perform patterning on engineered arrays allows the preparation of multiple-point nanopores on our substrate in a rapid and reproducible manner. The illustration of the process is reported in Fig. 4(c). TEM- and FIB-fabricated nanopores in MoS$_2$ flakes were characterized by means of TEM micrographs; Figure 4 reports examples of the obtained data from selected samples. As expected, while TEM sculpturing of a nanopore down to 2 nm can be easily achieved by controlling the duration of the exposure (Fig. 4(a) and (b)), in the case of FIB-milled holes, diameters just below 10 nm are the lower limit of this approach (Fig. 4(d)).

As described above, the holes in the MoS$_2$ layer are expected to result in vacancies in –S bonds that we use here as nucleation sites for the growth of Au nanoparticles (AuNPs). For the growth of AuNPs, a 2 mM HAuCl$_4$ solution was prepared in H$_2$O, with 20 μL dropped over the sculptured samples for different time durations to allow the AuNPs to grow. The dimensions of the obtained AuNPs depend on the duration of the deposition (see the SI for examples of different growths); AuNPs of approximately 5 nm in diameter were obtained using 30 seconds of incubation at room temperature and subsequent rinsing in H$_2$O. Under this condition, AuNPs were grown on both bare flakes and on flakes in which a nanohole was created. Figures 4(e) and 4(f) show TEM micrographs of MoS$_2$ decorated with AuNPs. The possibility of decorating a MoS$_2$ layer with metallic nanoparticles has been previously investigated in several recent papers[47-50], and it has also been demonstrated that the MoS$_2$ exposed edges are preferential nucleation sites. Consequently, in our case, it is possible to achieve AuNP growth in close proximity to the nanopore in almost all the cases; however, additional AuNPs can be present on the flakes. Control of the number of grown particles requires additional experimental optimizations. However, here, we are interested in a plasmonic phenomenon that we expect to observe with one or more AuNPs in close proximity to the pore.

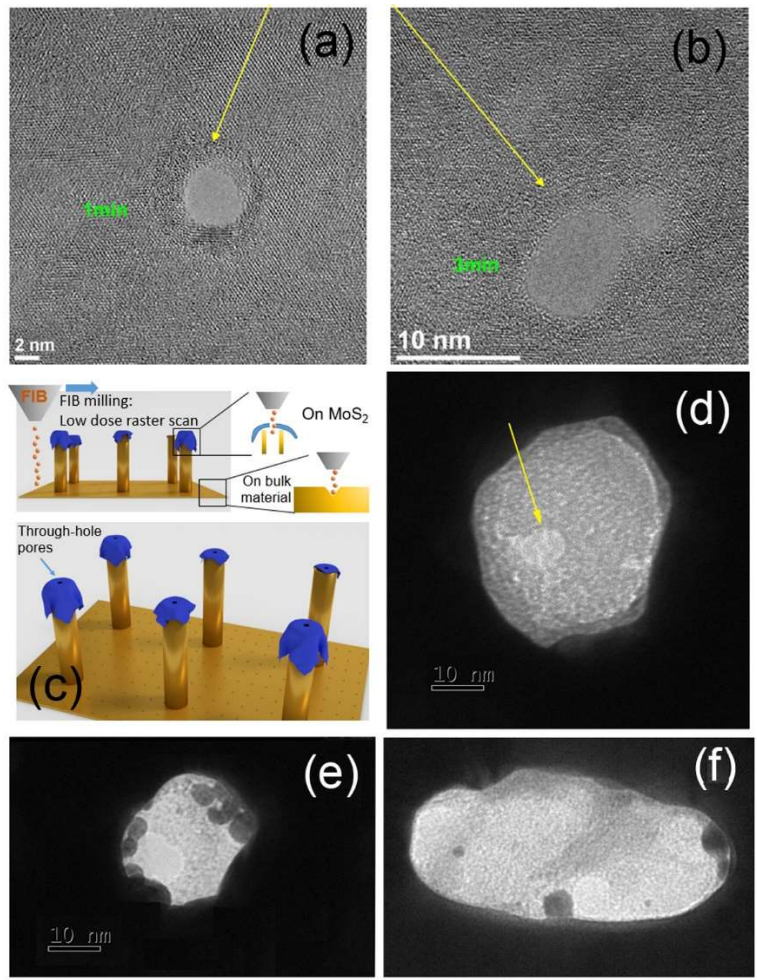

*Figure 4*. (a) TEM micrograph of a nanohole sculptured into a MoS$_2$ layer by means of HRTEM exposure for 1 minute; (b) TEM micrograph of a nanohole sculptured into MoS$_2$ layer by means of HRTEM exposure for 3 minutes; (c) illustration of the concept used for FIB milling of the MoS$_2$ nanopores array; (d) example of a nanopore prepared by means of FIB; (e)(f) example of AuNPs grown on a MoS$_2$ layer(s) after FIB milling.

Considering the simulations reported above and the proof-of-concept fabrication obtained, we expect a significant enhancement due to the presence of both the resonating antenna and AuNPs. To verify this enhancement, we performed Raman spectroscopy on our samples at two different wavelengths of excitation, i.e., 532 nm and 633 nm. The latter results appear to be resonant with the structure and are able to excite the mode at the antenna apex. The measurements were performed by using a Rainshaw InVia Microscope Raman system with a 50 × 0.95 NA objective, collecting the signal with a spectral resolution of 2.5 cm$^{-1}$ and an integration time of 1 second. The system was calibrated by using the intensity of the standard peak at 520 cm$^{-1}$ from a silicon substrate. Figure 5 reports the results of our measurements. In the top panel, a map over a large array of 256 3D antennas is reported. Raman shifts (excitation wavelength 532 nm) between 400 and 410 cm$^{-1}$ (in correspondence of A$_{1g}$ Raman mode) have been used to evaluate the coverage of the MoS$_2$ over the 256 points. According to the figure, the signal appears only in correspondence of the antennas. This result is a clear demonstration that the deposition strategy covers the desired elements in almost all the cases: over 85 % of the antennas are decorated with MoS$_2$ flakes. Figure 5(b) reports the statistical analysis over the 256 points with which we evaluate the number of layers corresponding to each MoS$_2$ flake. As illustrated in SI, the measured points were fitted by Lorentzian functions. The in-plane (E$^1_{2g}$ at ~ 380 cm$^{-1}$) and out-of-plane (A$_{1g}$ at ~ 404 cm$^{-1}$) Raman modes were always clearly visible and used in the analysis. The difference between the E$^1_{2g}$ and A$_{1g}$ modes ($\Delta f$) is known to steadily increase with the number of layers;[51-55] hence, this parameter can be a reliable quantity to count the number of layers of MoS$_2$. We used *Δf* to evaluate the percentage of single-layer flakes deposited on the considered array (Fig. 5(b), histogram). From our analysis, we can conclude that a single layer can be deposited over approximately 50 % of the antennas and nanopores. Regardless, we think that this can be improved acting on the exfoliation procedure to obtain better-quality, single-layer flakes in solution. Indeed, herein, we chose to follow a Li-intercalation protocol because of the clear advantages for our purposes with respect to other techniques. The resulting 1T-phase MoS$_2$ is well known to be more defective and, consequently, more reactive than the pristine, semiconductive hexagonal phase is.[17,56,57] This enhanced reactivity clearly promotes both anchoring the flake *via* thiol conjugation and nucleating AuNPs on MoS$_2$ defects. In addition, with respect to liquid-phase exfoliation, a Li-based method can provide stable suspensions without any surfactant (that may hamper both anchoring and plasmonic behavior) and with higher exfoliation degree,[58] which, in our case, was maximized by doubling the Li dose. Unfortunately, MoS$_2$ strongly tends to break up during the exfoliation, resulting in a quite large size dispersion (see SI). Even if our deposition procedure was proved to work with all nanosheet sizes, we believe that improving the synthetic procedure to have flakes with homogenous thickness and a controlled lateral dimension would allow further optimization of the deposition parameters and, consequently, enhancement of the performances of these hybrid systems. Finally, although our synthetic procedure is highly time consuming and low yielding, Li-exfoliation may be scaled-up by switching from chemical to electrochemical intercalation.[59,60] This process would allow the preparation of single-layer MoS$_2$ with higher throughput, which is necessary for the application of these types of systems on a large scale.

Finally, Raman spectroscopy has also been used to demonstrate, as a proof of concept, the resonance coupling between AuNPs grown on $MoS_2$ pores and the integrated plasmonic 3D antennas. Figure 5(c) reports the Raman shift collected with excitation wavelengths of 532 (see SI) and 633 nm in the presence of optimized 3D antennas (working at 633 nm) and with the addition of AuNPs. As seen in all the cases, both $E^1_{2g}$ and $A_{1g}$ Raman modes are observed. This observation is not surprising because, despite the use of 1T-$MoS_2$ flakes, it has been already demonstrated that $MoS_2$ may undergo a 1T→ 2H phase transition under laser beam.[61,62]

Although a detailed study of the Raman spectrum is far from the scope of this work, it is interesting to report that, using the excitation wavelength of 532 nm, a Lorentz function perfectly fits the experimental data and determines the two $E^1_{2g}$ and $A_{1g}$ Raman modes at 383 and 402 cm$^{-1}$, respectively, i.e., a $\Delta f$ below 20 cm$^{-1}$ equivalent to single-layer $MoS_2$. Moreover, this observation appears to be verified after AuNP growth, when a higher intensity in the Raman modes appears with a slightly increase in $\Delta f$ that we ascribe to the presence of the nanoparticles.

The presence of AuNPs induces a more significant enhancement in the collected spectra in the case of 633 nm excitation wavelength. This result can be caused by the combination of the additional enhancement from the plasmonic antennas and the resonant condition on excitation. In this latter case, it has not been possible to fit all the peaks with a single Lorentz function cause, as already reported[65], and additional modes appear at approximately 410 and 450 cm$^{-1}$.

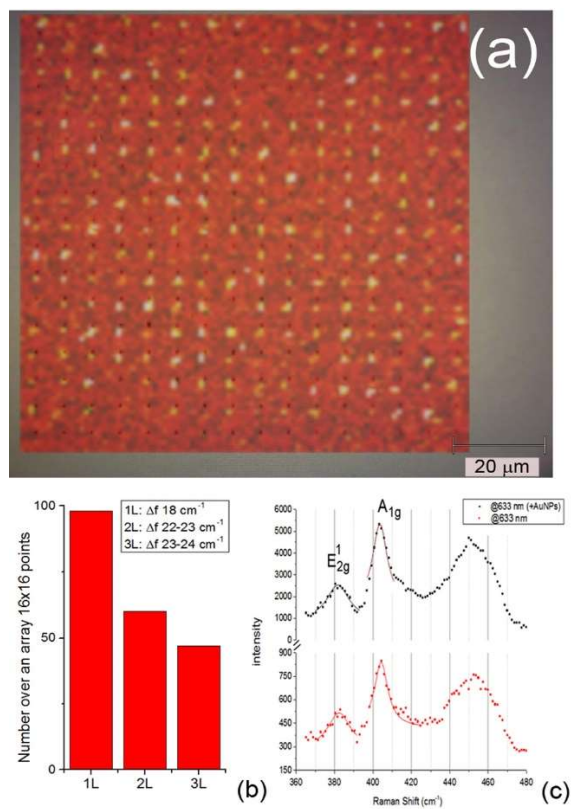

*Figure 5*. *Raman analysis on MoS$_2$ deposited on 3D metallic antennas. (a) Map over 256 antennas. Signal integrated between 400 and 410 cm$^{-1}$: 227 antennas give the expected signal. Note that the laser is slightly misaligned with respect to the optical image; (b) histogram reporting the statistical analysis on the number of layers based on Δf of the Raman modes; (c) Raman spectra of the same nanoantenna decorated with a MoS$_2$ flake before and after AuNP growth. Solid lines correspond to Lorentz curve fits, and dots correspond to experimental data.*

In summary, we presented a hybrid plasmonic 2D material structure able to generate a significant field confinement in close proximity of the nanopores. The fabrication procedure allows the preparation of ordered structures over large array using a low-cost procedure and without the use of complex lithographic processes. This strategy can be applied to not only MoS$_2$ but also many 2D materials for which the (always present) defects over the layer can be used to anchor the linker between the metallic nanopore and the flake. Moreover, the presence of defects and of edges on the 2D materials allows the nucleation of metallic nanoparticles, hence paving the way to integrate additional plasmonic elements over ordered hybrid structures.

We believe that such an architecture can be a key element for the realization of new hybrid devices for use in several applications, including photoluminescence, strong coupling and valley-polarization[64] studies, and single-molecule detection for DNA or protein sequencing. With respect to previously reported hybrid plasmonic nanostructures, our scheme significantly reduces the complexity of fabrication, leading to a more robust and low-cost approach for the integration of 2D materials with plasmonic nanopores.

AUTHOR INFORMATION

* Corresponding authors: Dr. Francesco De Angelis, Francesco.deangelis@iit.it

AUTHOR CONTRIBUTION

DG conceived the experiment and fabricated and characterized the structures; DM and SA prepared the $MoS_2$ flakes; EM proposed the functionalization protocol; NM performed the FEM simulations; MA and GG performed TEM microscopy; MDP helped with the data analysis; and FDA supervised the work.

ACKNOWLEDGMENTS

The research leading to these results has received funding from the European Research Council under the FET-Open: PROSEQO, Grant Agreement No. [687089].